\documentclass[thmsa,11pt]{article}
\usepackage{sw20lart}

\input tcilatex
\QQQ{Language}{
American English
}

\begin{document}

\author{\textbf{Gh. Zet }and\textbf{\ M. C. Neacsu} \\
Department of Physics\\
Technical University ''Gh. Asachi''\\
Iasi 6600, Romania}
\title{\textbf{A\ SELF-DUAL\ MODEL\ FOR\ THE SU(2) GAUGE THEORY}}
\date{}
\maketitle

\begin{abstract}
\ A model of SU(2) gauge theory in the space-time $R\times S^3$ is
constructed in terms of local gauge-invariant variables. A metric tensor $%
g_{\mu \nu }$ is defined starting with the components of the strength tensor 
$F_{\mu \nu }^k$ and of its dual $\widetilde{F}_{\mu \nu }^k$. It is shown
that the components $g_{\mu \nu }$ determine the gauge field equations if
some supplementary constraints are imposed. Two families of analitical
solutions of the field equations are also obtained.

\ 

\ 

\newpage\ 
\end{abstract}

\section{Introduction}

The gauge theory are usually formulated in terms of non-gauge-invariant
variables, like potentials $A_\mu ^k(x)$ \cite{1}. However, the physical
observables are gauge invariant. The connection between potentials and
observables is established by choosing a specific gauge. This rises many
difficulties even in the classical theories. For example, some solutions of
the field equations can be spherical symmetric in a chosen gauge, but they
may nor have this symmetry in other gauges. In addition, the potentials can
be singular in one and non-singular in others.

At the quantum level these difficulties are even more serious. For example,
the quantization of one and the same theory in different gauges can give
rise to non-equivalent theories.

However, the physicists prefer to use the non-gauge-invariant potentials
because of the following advantages:

(i) the theory has a local character;

(ii) the relativistic invariance is manifest;

(iii) the Lagrangian formalism applies in the standard form

Recently, some models of gauge theories on Euclidean and Minkowskian
3-dimensional spaces have been developed in terms of gauge-invariant
variables \cite{2}, \cite{3}. The fundamental quantity is the
gauge-invariant tensor $g_{ij}=-\frac 12Tr(^{*}F_i^{*}F_j)$, where $%
^{*}F_i=\frac 12\varepsilon _{ijk}F^{jk}$ is the dual of the gauge field
tensor $F^{ij}$. It has been shown that this metric satisfies the Einstein
equations with the right-hand side of a very simple form \cite{3}.

In a recent work \cite{4} we generalized this theory to the case of a curved
space-time. Namely, we developed a $SU(2)$ gauge theory on the
three-dimensional sphere $S^3$. The manifold $S^3$ is a space with constant
curvature, and the generalization of the theory to this case is not trivial.
We used the advantage that the dimensions of the $SU(2)$ group and of the $%
S^3$ sphere are the same.

In this paper, we develope a model of $SU(2)$ gauge theory in terms of local
gauge-invariant variables defined over a 4-dimensional space-time. We define
a metric tensor $g_{\mu \nu }$, $\mu ,\nu =0,1,2,3$, starting with the
components $F_{\mu \nu }^k$ and $\widetilde{F}_{\mu \nu }^k$ of the tensor
associated to the Yang-Mills field and, respectively, to its dual. The
components $g_{\mu \nu }$ are interpreted as new local gauge variables and
they are calculated for a particular gauge field defined over the $R\times
S^3$ space-time. It is shown that these components determine the gauge field
equations when some supplementary conditions are imposed.

Our metric $g_{\mu \nu }$ do not make the field strength self-dual. In order
to assure this property a convenient scale factor $\Delta $ is introduced in
the expression of the metric $g_{\mu \nu }.$ It is concluded that the
self-dual variables are nor compatible with the geometric structure of the
space-time $R\times S^3$.

Two families of analytical solutions for the field equations of the gauge
fields are also given. it is proven that these solutions are not of
solitonic type, but they can be used to determine the structure of vacuum
state for the gauge fields considered in this paper.

\section{ Gauge Potentials}

We consider the $SU(2)$ Yang-Mills theory in the space-time $R\times S^3$
endowed with the metric: 
\begin{equation}
ds^2=-dt^2+d\chi ^2+\sin ^2\chi (d\theta ^2+\sin ^2\theta d\varphi ^2) 
\tag{2.1}
\end{equation}
where $t$ is the time coordinate on the real line $R$ and $\chi ,\theta $
and $\varphi $ are the angular coordinates on the three-dimensional sphere $%
S^3$ \cite{5}. Let $P(M,G,\pi )$ be the principal fibre bundle with $%
M=R\times S^3$ as the base manifold and $G=SU(2)$ as the structural group.
The mapping $\pi :P\rightarrow M$ is the natural projection of $P$ onto $M$.
The infinitesimal generators of the $SU(2)$ group are chosen in the form: 
\begin{equation}
T_k=\frac 1{2i}\sigma _k\text{ , }k-1,2,3  \tag{2.2}
\end{equation}
where $\sigma _k$ are the Pauli matrices.

The corresponding structure equations are:

\begin{equation}
\left[ T_k,T_l\right] =\varepsilon _{klm}T_m  \tag{2.3}
\end{equation}
where $\varepsilon _{klm}$ is the antisymmetric tensor of rank 3 with $%
\varepsilon _{123}=1.$

The gauge potentials $A_\mu =A_\mu ^kT_k$ , with values in the Lie algebra
of the group $SU(2)$ , determine a connection on the principal fibre bundle $%
P(M,G,\pi )$ \cite{6}. We will consider the following ansatz \cite{7} for
the gauge potentials:

\begin{equation}
\begin{array}{cc}
A_0=-\Phi T_3, & A_1=0 \\ 
A_2=WT_2, & A_4=\cos \theta T_3-W\sin \theta T_1
\end{array}
\tag{2.4}
\end{equation}
where $\Phi $ and $W$ are two unknown functions depending only on the
variable $\chi $. Therefore, the components $A_\mu ^k$ of these gauge
potentials are the following:

\begin{equation}
\begin{array}{cccc}
A_o^1=0, & A_1^1=0, & A_2^1=0, & A_3^1=-W\cos \theta \\ 
A_0^2=0, & A_1^2=0, & A_2^2=W, & A_3^2=0 \\ 
A_0^3=-\Phi , & A_1^3=0, & A_2^3=0, & A_3^3=\cos \theta
\end{array}
\tag{2.5}
\end{equation}

The tensor of the gauge fields $F_{\mu \nu }=F_{\mu \nu }^kT_k$, with values
in the Lie algebra of $SU(2)$, is defined by the formula \cite{8}:

\begin{equation}
F_{\mu \nu }=\partial _\mu A_\nu -\partial _\nu A_\mu +[A_\mu ,A_\nu ] 
\tag{2.6}
\end{equation}
where $\partial _\mu $ denotes the derivatives with respect to the variables 
$x^\mu =(t,\chi ,\theta ,\varphi )$ and $\mu =0,1,2,3$.

The non-null components of this tensor are:

\begin{equation}
\begin{array}{cc}
F_{02}^1=W\Phi , & F_{13}^1=-W^{^{\prime }}\sin \theta , \\ 
F_{03}^2=\Phi W\sin \theta & F_{12}^2=W^{^{\prime }}, \\ 
F_{03}^3=\Phi ^{^{\prime }}, & F_{23}^3=(W^2-1)\sin \theta
\end{array}
\tag{2.7}
\end{equation}
where $\Phi ^{^{\prime }}=\frac{d\Phi }{d\chi }$ and $W^{^{\prime }}=\frac{dW%
}{d\chi }$ are the derivatives of the two functions $\Phi $ and $W$ with
respect to the variable $\chi $.

In the next section we will determine a metric tensor $g_{\mu \nu }$
starting with the tensor $F_{\mu \nu }$ and its dual $\widetilde{F}_{\mu \nu
}^{}$. The components of this metric tensor will be interpreted as a local
gauge-invariant variables for the $SU(2)$ Yang-Mills theory on the
space-time $R\times S^3$.

\section{Local gauge-invariant variables}

The gauge potentials $A_\mu ^k$ are not invariant under the gauge
transformations. In order to obtain new variables, which are invariant, we
define first the dual $\widetilde{F}_{\mu \nu }^{}$ of the strength tensor
field $F_{\mu \nu }$ \cite{9}:

\begin{equation}
F_{\mu \nu }=\frac 12\varepsilon _{\mu \nu \sigma \tau }\widetilde{F_{\mu
\nu }^{}}  \tag{3.1}
\end{equation}

where $\varepsilon _{\mu \nu \sigma \tau }$ is the antisymmetric tensor of
rank 4, with $\varepsilon _{0123}=1$. Then, using the above result (2.7), we
obtain the following non-null components of the dual $\widetilde{F}_{\mu \nu
}^{}$:

\begin{equation}
\begin{array}{cc}
\widetilde{F_{02}^1}=W^{^{\prime }}\sin \theta , & \widetilde{F_{02}^1}%
=-W\Phi , \\ 
\widetilde{F_{12}^2}=\Phi W\sin \theta , & \widetilde{F_{03}^2}=W^{^{\prime
}}, \\ 
\widetilde{F_{01}^3}=(W^2-1)\sin \theta & \widetilde{F_{23}^3}=\Phi
^{^{\prime }}
\end{array}
\tag{3.2}
\end{equation}

Now, we introduce new local gauge-invariant variables $g_{\mu \nu }$, given
by \cite{10}:

\begin{equation}
g_{\mu \nu }=\frac 1{3\Delta ^{1/3}}\varepsilon _{}^{klm}F_{\mu \alpha }^k%
\widetilde{F_{\alpha \beta }^l}F_{\beta \nu }^m  \tag{3.3}
\end{equation}
and:

\begin{equation}
g_{}^{\mu \nu }=\frac 2{3\Delta ^{1/3}}\varepsilon _{}^{klm}\widetilde{%
F_{\mu \alpha }^k}F_{\alpha \beta }^l\widetilde{F_{\beta \nu }^m}  \tag{3.4}
\end{equation}
Here, $\Delta $ is a scale factor which will be chosen of a convenient form
in what follows.

Introducing the expressions (2.7) and (3.2) of the tensor $F_{\mu \nu }$ and
respectively of its dual $\widetilde{F}_{\mu \nu }^{}$, we obtain the
following non-null components of $g_{\mu \nu }$: 
\begin{equation}
\begin{array}{c}
g_{00}=\frac 2{\Delta ^{1/3}}W^2\Phi ^2\Phi ^{^{\prime }}\sin \theta , \\ 
g_{11}=\frac 2{\Delta ^{1/3}}W^{\prime 2}\Phi ^{^{\prime }}\sin \theta , \\ 
g_{22}=\frac 2{\Delta ^{1/3}}WW^{^{\prime }}\Phi (W^2-1)\sin \theta , \\ 
g_{33}=\frac 2{\Delta ^{1/3}}WW^{^{\prime }}\Phi (W^2-1)\sin ^3\theta ,
\end{array}
\tag{3.5}
\end{equation}
Having these quantities determined, we introduce a new metric manifold,
whose line element written in the variables $(t,\chi ,\theta ,\varphi )$ is:

\begin{equation}
d\sigma ^2=g_{00}dt^2+g_{11}d\chi ^2+g_{22}d\theta ^2+g_{33}d\varphi ^2 
\tag{3.6}
\end{equation}
or: 
\begin{equation}
d\sigma ^2=-\frac{2W^2\Phi ^2\Phi ^{^{\prime }}\sin \theta }{\Delta ^{1/3}}%
\left[ -dt^2-\frac{W^{\prime 2}}{W^2\Phi ^2}d\chi ^2+\frac{W^{^{\prime
}}(1-W^2)}{W\Phi \Phi ^{^{\prime }}}\left( d\theta ^2+\sin ^2\theta d\varphi
^2\right) \right]  \tag{3.7}
\end{equation}

If we chose now the scale factor $\Delta $ in the form: 
\begin{equation}
\Delta ^{1/3}=-2W^2\Phi ^2\Phi ^{^{\prime }}\sin \theta  \tag{3.8}
\end{equation}
then (3.7) reduces to: 
\begin{equation}
d\sigma ^2=-dt^2-\frac{W^{\prime 2}}{W^2\Phi ^2}d\chi ^2+\frac{W^{^{\prime
}}(1-W^2)}{W\Phi \Phi ^{^{\prime }}}\left( d\theta ^2+\sin ^2\theta d\varphi
^2\right)  \tag{3.9}
\end{equation}

The line element (3.9) coincides with that of the space-time $R\times S^3$
given in (2.1) if we impose the following supplementary conditions:

\begin{equation}
\begin{array}{cc}
W^{^{\prime }}=\frac{dW}{d\chi }=iW\Phi , & 
\end{array}
\Phi ^{^{\prime }}=\frac{d\Phi }{d\chi }=i\frac{1-W^2}{\sin ^2\chi } 
\tag{3.10}
\end{equation}

But the conditions (3.10) are nothing than the field Yang-Mills equations
for the potentials $A_\mu ^k$ \cite{9}. In fact, the Yang-Mills equations
are differential equations of second order; in the case of the ansatz
(2.40), they have the following form: 
\begin{equation}
\frac d{d\chi }\left( \frac{d\Phi }{d\chi }\sin ^2\chi \right) =2W^2\Phi 
\tag{3.11}
\end{equation}
\begin{equation}
\frac d{d\chi }\left( \frac{dW}{d\chi }\right) =-W\Phi ^2-\frac{(1-W^2)}{%
\sin ^2\chi }W  \tag{3.12}
\end{equation}

It is easy to verify that the equations (3.11)-(3.12) are equivalent with
the first order equations given in (3.10).

Therefore, we conclude that the scale factor $\Delta $ chosen in (3.8)
together with the field equations of the gauge potentials (2.5), reduce the
metric $g_{\mu \nu }$ to that of the space-time $R\times S^3$.

It is important to remark that choosing $\Delta $ as in (3.8) we do not
obtain a self-dual gauge theory, i.e. $F_{\mu \nu }$ is not a self-dual
tensor. This means that the following condition of self-duality is not
satisfied \cite{10}: 
\begin{equation}
\frac 1{2\sqrt{g}}\varepsilon ^{\alpha \beta \sigma \tau }g_{\mu \alpha
}g_{\nu \beta }F_{\sigma \tau }=F_{\mu \nu }  \tag{3.13}
\end{equation}
where $g=\det (g_{\mu \nu }).$ Therefore, the metric (2.1) of the space-time
is not compatible with a self-dual $SU(2)$ gauge theory.

In order to satisfy the condition (3.13) we should choose the factor $\Delta 
$ in the form: 
\begin{equation}
\Delta =8\Phi ^2W^2W^{^{\prime }2}(W^2-1)\sin ^3\theta  \tag{3.14}
\end{equation}

In this case we obtain: 
\begin{equation}
g=\det (g_{\mu \nu })=\frac 14\Delta ^{2/3}  \tag{3.15}
\end{equation}
However, the metric components $g_{\mu \nu }$ in (3.5) do not correspond to
the geometry of the space-time $R\times S^3$ when $\Delta $ is given by
(3.14). Therefore, the space-time $R\times S^3$ do not admit a self-dual $%
SU(2)$ gauge theory.

\section{Solutions of the field equations}

We have obtained two families of analytic solutions of the field equations
(3.10) \cite{9}. First of them is:

\begin{equation}
\Phi (\chi )=i\mu \cot (\mu \chi )-i\cot (\chi )  \tag{4.1}
\end{equation}

\begin{equation}
W(\chi )=-\frac{\mu \sin \chi }{\sin (\mu \chi )}  \tag{4.2}
\end{equation}
where $\mu $ is an arbitrary constant. The Chern index of this solution is:

\begin{equation}
c_2=\int_0^\pi \left[ \Phi ^{^{\prime }}(1-W^2)-2WW^{^{\prime }}\Phi \right]
=\Phi (1-W^2)\mid _0^\pi =0  \tag{4.3}
\end{equation}
This value of $c_2$ shows that the solutions (4.1)-(4.2) in first family are
not of instanton type.

The second family of solutions for the field equations 93.10) is given by 
\cite{9}:

\begin{equation}
\Phi (\chi )=i\nu \coth (\mu \chi )-i\cot (\chi )  \tag{4.1}
\end{equation}

\begin{equation}
W(\chi )=-\frac{\nu \sin \chi }{\sinh (\nu \chi )}  \tag{4.2}
\end{equation}
where $\nu $ is an arbitrary real constant. the Chern index has, in this
case, the value $c_2=\infty $. Therefore, the solutions in this family are
not of instanton type too.

However, the solutions in the two families can be used for studying the
vacuum structure of the fields under considerations.

\section{ Concluding remarks}

In this paper we have developed a gauge theory in terms of local
gauge-invariant variables defined over the space-time $R\times S^3$. A
metric tensor $g_{\mu \nu }$, $\mu ,\nu =0,1,2,3$, has been constructed
starting with the components of the strength tensor $F_{\mu \nu }$ and of
its dual $\widetilde{F}_{\mu \nu }$. The components $g_{\mu \nu }$ have been
interpreted as new local gauge-invariant variables and they have been
calculated for some particular gauge fields defined over the space-time $%
R\times S^3$. We proved that these components determine the gauge field
equations if some supplementary constraints are imposed. On the other hand
we showed that the metric tensor $g_{\mu \nu }$ do not make the field
strength $F_{\mu \nu }$ self-dual. In order to assure this property, a
convenient scale factor $\Delta $ has been introduced in the expression of $%
g_{\mu \nu }$. We concluded that the self-dual variables are not compatible
with the geometric structure of the space-time $R\times S^3$.

Finally, two families of analitical solutions of the gauge field equations
have been obtained. we proved that these solutions are not of instanton
type, but they could be used to determine the structure of the vacuum state
for the gauge fields considered in this paper.

\end{document}